\documentclass{article}

\newcommand{\mq}{\mathbf{q}}
\newcommand{\mpsi}{\boldsymbol{\psi}}
\newcommand{\mbpsi}{\bar{\boldsymbol{\psi}}}

\newcommand{\mLambda}{\boldsymbol{\Lambda}}

\newcommand{\mJ}{\boldsymbol{J}}
\newcommand{\mbJ}{\overline{\boldsymbol{J}}}

\newcommand{\p}[1]{(\ref{#1})}

\newcommand{\cF}{{\cal F}}
\newcommand{\cE}{{\cal E}}

\newcommand{\cD}{{\cal D}}

\newcommand{\cG}{{\cal G}}
\newcommand{\cA}{{\cal A}}

\newcommand{\bT}{{\overline T}}

\newcommand{\bD}{{\overline D}}

\newcommand{\bQ}{{\overline Q}}
\newcommand{\bS}{{\overline S}}
\newcommand{\bW}{{\overline W}}
\newcommand{\bK}{{\overline K}}
\newcommand{\bJ}{{\overline J}{}}

\newcommand{\bpsi}{{\bar\psi}{}}

\newcommand{\bnabla}{{\overline \nabla}}

\newcommand{\be}{\begin{equation}}
\newcommand{\ee}{\end{equation}}
\newcommand{\bea}{\begin{eqnarray}}
\newcommand{\eea}{\end{eqnarray}}

\newcommand{\ba}{\begin{array}} \newcommand{\ea}{\end{array}}

\def\im{{\rm i}}

\newcommand{\nn}{\nonumber}

\usepackage{amscd,amsmath,amssymb}

\def\theequation{\arabic{section}.\arabic{equation}}

\begin{document}
\thispagestyle{empty}
\vspace{2cm}
\begin{flushright}
\end{flushright}\vspace{2cm}
\begin{center}
{\Large\bf Component on-shell actions of supersymmetric 3-branes\\[1cm]
 II. 3-brane in  D=8}
\end{center}
\vspace{1cm}

\begin{center}
{\large\bf S.~Bellucci${}^a$, N.~Kozyrev${}^b$, S.~Krivonos${}^{b}$ and A.~Sutulin${}^{a,b}$}
\end{center}

\begin{center}
${}^a$ {\it
INFN-Laboratori Nazionali di Frascati,
Via E. Fermi 40, 00044 Frascati, Italy} \vspace{0.2cm}

${}^b$ {\it
Bogoliubov  Laboratory of Theoretical Physics, JINR,
141980 Dubna, Russia} \vspace{0.2cm}

\end{center}
\vspace{2cm}

\begin{abstract}\noindent
In the present paper we explicitly construct the on-shell supersymmetric component action for a
3-brane moving in $D=8$ within the nonlinear realizations framework. Similarly to the previously considered case of the super 3-brane in $D=6$,  all ingredients
entering the component action follow from the nonlinear realizations approach. The component action of the 3-brane possesses $N=4, d=4$ supersymmetry partially broken
to $N=2, d=4$ one. The basic Goldstone superfield is the generalized version of $N=2,d=4$ hypermultiplet. The action has a structure, such that  all terms of higher
orders in the fermions are hidden inside the covariant derivatives and vielbeins. The main part of the component action mimics its bosonic cousin in which the
ordinary space-time derivatives and the bosonic worldvolume are replaced by their covariant (with respect to broken supersymmetry) supersymmetric analogs.
The spontaneously broken supersymmetry fixes the Ansatz for the component action, up to two constant parameters.
The role of the unbroken supersymmetry is just to fix these parameters.
\end{abstract}

\newpage
\setcounter{page}{1}
\setcounter{equation}{0}
\section{Introduction}
In the famous super p-brane scan \cite{branescan} there are only two branes with four-dimensional worldvolume - 3-brane in $D=6$ and in $D=8$.
While the first 3-brane has been considered in \cite{d3d6} and then further analysed  in many other approaches including the superfield
ones \cite{bg1,rt,r}, for the best of our knowledge the 3-brane in $D=8$ has been not considered yet. It is a bit strange, because
on the worldvolume the effective action for such 3-bane will be just the action for $N=2, d=4$ hypermultiplet \cite{hyper} -
the superfield which probably is not less known than the $N=2, d=4$ vector supermultiplet. Such an action has to possess hidden,
spontaneously broken $N=2$ supersymmetry which,
together with the manifest $N=2,d=4$ one, forms a $N=4, d=4$ supersymmetry algebra with four central charges. Such a situation,
probably, can be explained by the on-shell
nature of the hypermultiplet constraints in the standard superspace, so the using of harmonic \cite{HSS} or/and projective
\cite{PSS} superspaces\footnote{See also \cite{sk3} and references therein.} is unavoidable.  Another
reason probably is related with the fact that, after dimensional reduction from the known actions in higher dimensions
and fixing the $\kappa$-supersymmetry, we will end up
with a long tail of fermionic terms having no geometric meaning. Moreover, in the theories with partially broken supersymmetry there is the possibility of redefining the
fermionic components in many different ways: starting from the fermions of the linear realization and finishing by the fermions of the nonlinear realization. Under such redefinitions
the action changes drastically and {\it a priori} it is unknown which basis is preferable.

In this second part of our paper we construct and analyse in details the action of supersymmetric 3-brane in $D=8$ using the nonlinear realization approach \cite{nlr1, nlr2}
properly modified for the construction of component actions in \cite{BIKS1, BKSu1}. In our approach we are paying  much attention to the broken
supersymmetry, almost ignoring the unbroken one. Thanks to the fact that all physical components appear as the parameters of the
corresponding coset, all of them have Goldstone nature.
Keeping in mind that the Goldstone fermions, accompanying the partial breaking of supersymmetry,  can enter the component action
either through the covariant derivatives or vielbeins, only, the Ansatz for the action having a proper bosonic limit contains
just two constants (one constant is related with the possibility to add Wess-Zumino term to the action). The fixing of these
constants is the role playing by unbroken supersymmetry.

The plan of our paper goes as follows. In Section 2 we provide the superspace description of the 3-brane in $D=8$.
The constraints on the basic $N=2, d=4$ superfield (covariantized hypermultiplet conditions) and the equations of motion - that
is all that we can get in this way. Section 3 is the main part of our paper. There we derive the bosonic action
of our 3-brane, propose the Ansatz for the component action, fixing the parameters entering into such an Ansatz and, finally,
demonstrate the relations between spinor derivatives of the fermionic superfields and space-time derivatives of the hypermultiplet.
We conclude with a short discussion and further perspectives of our approach. The Appendix contains
purely technical material, concerning the evaluation of the Cartan forms and covariant derivatives.
\setcounter{equation}{0}
\section{Three-brane in $D=8$}
The action of the $N=2$ supersymmetric three-brane in $D=8$ is a minimal action for the
Goldstone $N=2, d=4$ hypermultiplet accompanying the spontaneous breaking of $N=1, D=8$ supersymmetry down to $N=2, d=4$ one, or, in other words, the breaking
of $N=4$ supersymmetry to $N=2$ in four dimensions \cite{branescan}. Despite the fact that  this information should be
enough to construct the superfield action of the three-brane in terms of the $N=2, d=4$ hypermultiplet, for the best of
our knowledge, such an action has been not constructed yet. Our goal in this paper is to carry out a less ambitious task -
to construct the on-shell component action for such a system. In our construction we are going to use the building blocks having clear geometric properties with respect
to broken $N=2$ supersymmetry.  The basic tool for this is, similarly to the first part of this paper \cite{part1},
the method of nonlinear realizations \cite{nlr1,nlr2} adopted for this task in \cite{BIKS1, BKSu1}. Our procedure includes three steps:
a) construction of the superfield equations of motion, b) deriving the  bosonic action, c) construction of the full component action.

\subsection{Superfield equations of motion}
From the $d=4$ standpoint the $N=1, D=8$ supersymmetry algebra is a four central charges extended $N=4, d=4$
Poincar\'e superalgebra with the following basic relations:
\bea\label{superP}
&& \big\{ Q_{\alpha}^i,  \bQ_{j \dot\alpha}   \big\}  =  2 \delta^i_j \left( \sigma^A \right)_{\alpha\dot\alpha} P_A, \qquad
\big\{ S_{\alpha}^a, \bS_{b \dot\alpha}   \big\} =  2 \delta^a_b \big( \sigma^A \big)_{\alpha\dot\alpha} P_A, \nn \\
&&\big\{ Q_{\alpha}^i,  S_{\beta}^a   \big\}  =  2 \epsilon_{\alpha\beta} Z^{ia},\qquad
\big\{  \bQ_{i \dot\alpha}, \bS_{a \dot\beta} \big\}  =  2\epsilon_{\dot\alpha\dot\beta} Z_{ia}.
\eea
As a reminder about its eight-dimensional nature, the superalgebra \p{superP} possesses the $so(1,7)$ automorphism algebra.
Again, from $d=4$ perspective, $so(1,7)$ algebra contains  the $d=4$ Lorentz algebra $so(1,3)$ generated by  $L_{AB}$, $su(2)\oplus su(2)$
subalgebra with generators $T^{ij}$ and $R^{ab}$, respectively, and the generators $K^{ia}_A$ from the coset $SO(1,7)/SO(1,3)\times SU(2)\times SU(2)$.
The full set of commutation relations can be found in Appendix.

Keeping the $d=4$ Lorentz and $SU(2)\times SU(2)$ symmetries linearly realized, we will choose the coset element as
\be\label{4dcoset}
g = e^{\im x^A P_A} e^{\theta^\alpha_i Q^i_\alpha + \bar\theta^{i \dot\alpha}\bQ_{i \dot\alpha}} e^{\im \mq^{ia}Z_{ia}} e^{\mpsi^{\alpha}_a S^a_\alpha
+ \mbpsi^{a\dot\alpha} \bS_{a\dot\alpha}} e^{\im \mLambda^A_{ia} K_A^{ia}}.
\ee
Here, we associated the $N=2, d=4$ superspace coordinates $x^A, \theta_i ^\alpha, \bar\theta{}^{i \dot\alpha}$ with the generators
$P_A, Q^i_\alpha, \bQ_{i \dot\alpha}$ of unbroken $N=2$ supersymmetry. The remaining coset parameters are Goldstone $N=2, d=4$ superfields.

The transformation properties of the coordinates and superfields with respect to all symmetries can be found by acting from the left on the coset element
$g$ \p{4dcoset} by the different elements of $N=4, d=4$ central charges extended Poincar\'e supergroup. In particular, for the unbroken
$(Q,\bQ)$ and broken $(S,\bS)$ supersymmetries we have
\begin{itemize}
\item Unbroken supersymmetry:
\be\label{susyQ}
\delta_Q x^A = \im \left( \epsilon^\alpha_i \bar\theta{}^{i \dot\alpha}+ \bar\epsilon{}^{\,i \dot\alpha} \theta_i^\alpha\right)\left( \sigma^A\right)_{\alpha\dot\alpha}, \quad
\delta_Q\theta_i^\alpha=\epsilon_i^\alpha, \quad \delta_Q\bar\theta{}^{i \dot\alpha}=\bar\epsilon{}^{\,i \dot\alpha};
\ee
\item Broken supersymmetry:
\be\label{susyS}
\delta_S x^A = \im \big( \varepsilon_a^{\alpha}\mbpsi^{a \dot\alpha} + \bar\varepsilon^{\,a \dot\alpha}\mpsi_a^{\alpha} \big)\left( \sigma^A  \right)_{\alpha\dot\alpha},\quad
\delta_S \mpsi_a^\alpha = \varepsilon_a^\alpha, \quad \delta_S\mbpsi^{a \dot\alpha} = \bar\varepsilon^{\,a \dot\alpha}, \quad
\delta_S \mq^{i a} = 2\im \left( \varepsilon^a_{\alpha} \theta^{i \alpha}+ \bar\varepsilon^{\,a}_{\dot\alpha} \bar\theta^{i \dot\alpha}\right).
\ee
\end{itemize}
The local geometric properties of the system are specified by the Cartan forms. The purely technical calculations of these forms,
semi-covariant derivatives and their algebra are summarized in Appendix.

As we already demonstrated in \cite{BIKS1, BKSu1, part1}, the covariant superfield equations of motion may be obtained by imposing
the following constraints on the Cartan forms:
\be\label{dyn}
\omega_Z = {\bar\omega}_Z=0 \;\; (a)\,, \qquad \left.  \omega_S\right|=\left . {\bar\omega}_S\right| =0 \;\;(b)\,,
\ee
where $|$ means the $d\theta$- and $d\bar\theta$-projections of the forms. These constraints are similar to superembedding
conditions (see e.g. \cite{Dima1} and references therein).

The constraints (\ref{dyn}a)  result in the following equations:
\bea
&& \nabla_\alpha^j \mq_{ia} + 2\im \mpsi_{a\alpha}\delta^j_i =0, \quad
\bnabla_{j\dot\alpha}\mq_{ia} +2\im \mbpsi_{a\dot\alpha}\epsilon_{ij}=0, \label{eq1}\\
&& \nabla_A \mq^{ia} =  \mLambda_{A}^{jb} \frac{\tanh\sqrt{2\mLambda^B_{jb} \mLambda^{ia}_B}}{\sqrt{2\mLambda^B_{jb} \mLambda^{ia}_B}}\,.\label{eq2}
\eea
These equations allow us to express the superfields $\mpsi_a^\alpha, \mbpsi^{a \dot\alpha}$ and $\mLambda^{ia}_A$ through the covariant
derivatives of superfields $\mq^{ia}$  (this is the so called Inverse Higgs phenomenon \cite{IH}):
\bea
&& \mpsi_{a\alpha} = \frac{\im}{4}\, \nabla^k_\alpha \mq_{ka}, \quad \mbpsi_{a\dot\alpha} = \frac{\im}{4}\, \bnabla_{k\dot\alpha}\mq^k_a, \label{eq1a} \\
&& \mLambda_{A}^{ia} = \nabla_A \mq^{ia}+ \ldots , \label{eq2a}
\eea
where in \p{eq2a} we explicitly write only the leading, linear in $\nabla_A \mq^{ia}$ term.
In addition, from \p{eq1} it follows that
\be\label{hyper}
\nabla_\alpha^{(i} \mq_a^{j)}=0, \quad \bnabla_{\dot\alpha}{}^{(i} \mq^{j)}_a=0.
\ee
Clearly, these are just a covariantized version of the hypermultiplet conditions \cite{hyper} which put theory on-shell.

The constraints (\ref{dyn}b) follow from (\ref{dyn}a), but their explicit form helps to simplify the consideration.  Firstly, the $d\bar\theta$ ($d\theta$) projection of the form $\omega_S$ ($\bar\omega_S$) relates the spinor
derivative of the superfields $\mpsi^\alpha_a, \mbpsi^{a\dot\alpha}$ and $x-$derivative of the superfield $\mq^{ia}$
\bea
&& \bnabla_{k\dot\gamma} \mpsi^\alpha_b \equiv \big(\, \mbJ{}^\alpha_{\dot\gamma} \big)_{kb}
=  \mLambda^\gamma_{ck\dot\gamma} \bigg( \frac{ \tanh \sqrt{T}}{\sqrt{T}} \,\bigg)_{b\gamma}^{c\alpha}=\mLambda_{k b \gamma}^\alpha+\ldots, \nn \\
&& \nabla^k_\gamma \mbpsi^{b\dot\alpha}\equiv \big(\,\mJ_\gamma^{\dot\alpha} \big)^{kb}
= \mLambda_\gamma^{kc\dot\gamma}\bigg( \frac{ \tanh \sqrt{\bT}}{\sqrt{\bT}} \,\bigg)_{c\dot\gamma}^{b\dot\alpha}=\mLambda_\gamma^{k b \dot\alpha}+\ldots. \label{eq3}
\eea
At the same time,  the $d\theta$ ($d\bar\theta$) projection of the form $\omega_S$ ($\bar\omega_S$) gives the equations
\be\label{eq4}
\nabla^i_{\alpha} \mpsi_a^\beta =0, \qquad \bnabla_{i \dot\alpha} \mbpsi^{a \dot\beta}= 0.
\ee
The invariance of the equations \p{eq1}, \p{eq2}, \p{eq3} and \p{eq4} with respect to $N=2,d=4$ superalgebra \p{D8Poincare}, \p{superD8Poincare}
is guaranteed by the invariance of the conditions \p{dyn}.

Thus, we conclude that the supersymmetric 3-brane in $D=8$ is described by the covariantized $N=2,d=4$ hypermultiplet. Unfortunately, it is not so simple to write down even the
bosonic equations of motion which follow from \p{hyper}. In the next Section we obtain the proper bosonic action starting from its invariance with respect
to the bosonic subalgebra  \p{D8Poincare}, and then  we will construct the full on-shell component action for our supersymmetric 3-brane.

\setcounter{equation}{0}
\section{Component action}
As we already noted, it is not clear how to construct the superfield action within the nonlinear realizations approach.
It is even technically hard to extract the bosonic equations of motion from the superfield ones \p{hyper}. Therefore, we are going
to construct the on-shell supersymmetric component action within the nonlinear realizations approach following the approach developed in \cite{BIKS1, BKSu1, part1}.
The useful ingredients for this construction include the reduced Cartan forms and reduced
covariant derivatives, covariant with respect to broken supersymmetry only. The basic steps of our approach are
\begin{itemize}
\item construction of the bosonic action
\item covariantization of the bosonic action with respect to broken supersymmetry
\item construction of the Wess-Zumino terms
\item imposing the invariance with respect to the unbroken supersymmetry.
\end{itemize}
Let us perform all these steps for the supersymmetric 3-brane in $D=8$.
\subsection{Bosonic action}
In principle, the bosonic equations of motion can be extracted from the superfield equations \p{hyper}. But the calculations are rather involved, because
one has to express the superfields $\mLambda_A^{ia}$ in terms of $\nabla_A \mq^{ia}$ (see eq. \p{eq2}). Instead, one can construct
the corresponding action directly, using the fact that such an action should possess invariance with respect to $D=8$ Poincar\'{e} symmetry spontaneously broken to
$d=4$. One of the key ingredients of such a construction is the bosonic limit of the Cartan forms \p{formsA} which explicitly reads
\bea\label{bos_forms1}
\left(\omega_P \right)^A_{bos} &=& dx^B \cosh \sqrt{2 \Lambda_B^{jb} \Lambda_{jb}^A}
-2 d q^{jb} \Lambda^C_{jb}\, \frac{\sinh\sqrt{2\Lambda_C^{jb} \Lambda_{jb}^A}}{\sqrt{2\Lambda_C^{jb} \Lambda_{jb}^A}}\,, \nn \\
\left(\omega_Z\right)^{ia}_{bos} &=& d q^{jb} \cosh \sqrt{2 \Lambda_{jb}^A \Lambda_A^{ia}}
- d x^A  \Lambda_{A}^{jb}\, \frac{\sinh\sqrt{2\Lambda^B_{jb} \Lambda^{ia}_B}}{\sqrt{2\Lambda^B_{jb} \Lambda^{ia}_B}}\,.
\eea
The bosonic part of our constraints \p{dyn}
$$ \left(\omega_Z\right)_{bos} =\left(\bar\omega_Z\right)_{bos}=0,$$
will result in the bosonic analog of the relations \p{eq2}
\be\label{eq2bos}
\partial_A q^{ia} =  \Lambda_{A}^{jb}\, \frac{\tanh\sqrt{2\Lambda^B_{jb} \Lambda^{ia}_B}}{\sqrt{2\Lambda^B_{jb} \Lambda^{ia}_B}}\,.
\ee
Plugging these expressions in the form $\left(\omega_P \right)^A_{bos}$ \p{bos_forms1} one may obtain
\be\label{volumef1}
\left(\omega_P\right)^A_{bos} = dx^B\; e_B{}^A = dx^B \left( \frac{1}{\cosh\sqrt{2 \Lambda^B_{jb} \Lambda^{jb}_A}} \right).
\ee
Now, the unique invariant which can be constructed from the forms $\left(\omega_P\right)^A_{bos}$ is a volume form which explicitly reads $d^4x \det (e)$.
Thus, the invariant bosonic action is uniquely defined to be
\be\label{bosac1}
S_{bos} = \int d^4x \det (e).
\ee
Using the explicit expressions \p{eq2bos} and \p{volumef1}, one may find the simple expression for the "metric" $g_{AB}$ in terms of $\partial_A q^{ia}$
\be\label{metric}
g_{AB} \equiv e_A^C e_{CB} =\eta_{AB} - 2\partial_{A} q^{ia}\partial_B q_{ia}, \quad g = \det g_{AB}\,,
\ee
and, therefore, the bosonic action acquires the form
\be\label{bos_action}
S_{bos} = \int d^4 x \; \sqrt{- g}\;.
\ee
This is the static gauge Nambu-Goto action for the 3-brane in $D=8$. One may explicitly check that the action \p{bos_action} is invariant with respect
to $K^{ia}_A$ transformations (with the parameter $\cA^A_{ia}$)  from the coset $SO(1,7)/SO(1,3)\times SU(2)\times SU(2)$ realized as
\be\label{trans1}
\delta_K x^A = 2 \cA^A_{ia} q^{ia}, \quad \delta_K q^{ia} =\cA^{ia}_A x^A.
\ee
and therefore, it is invariant with respect to the whole $D=8$ Poincar\'{e} group.

\subsection{Covariantization  with respect to broken supersymmetry}
Working in the component approach, we cannot straightforwardly construct the Ansatz for the action which possesses the unbroken supersymmetry. In contrast,
the broken $(S, \bS)$ supersymmetry can be maintained quite easily due to the transformations $\delta_S \theta^\alpha_i=\delta_S \bar\theta{}^{i\dot\alpha}=0$.
Thus, the first task is to modify the bosonic action \p{bos_action} in such a way as to achieve invariance with respect to broken supersymmetry.
Due to the transformation laws \p{susyS}, the coordinates $x^A$ and the first components $(q^{ia}, \psi_a^\alpha, \bpsi^{a \dot\alpha})$ of the superfields
$(\mq^{ia}, \mpsi_a^\alpha, \mbpsi^{a \dot\alpha})$ transform under broken supersymmetry as follows:
\be\label{susyS1}
\delta_S x^A = \im \left( \varepsilon_a^{\alpha}\bpsi^{a \dot\alpha} + \bar\varepsilon^{\,a \dot\alpha}\psi_a^{\alpha}   \right)
\left( \sigma^A  \right)_{\alpha\dot\alpha},\quad \delta_S \psi_a^\alpha = \varepsilon_a^\alpha, \quad \delta_S\bpsi^{a \dot\alpha} = \bar\varepsilon^{\,a \dot\alpha},
\quad \delta_S q^{i a} = 0.
\ee
Thus, the volume $d^4x$ and the derivatives $\partial_A q^{ia}$ are not the covariant objects. In order to find the proper objects, let us consider the reduced coset
element \p{4dcoset}
\be\label{4dcoset1}
g_{red} = e^{\im x^A P_A}  e^{\im q^{ia}Z_{ia}} e^{\psi^{\alpha}_a S^a_\alpha + \bpsi^{a\dot\alpha} \bS_{a\dot\alpha}},
\ee
where the fields $(q^{ia}, \psi^{\alpha}_a, \bpsi^{a\dot\alpha})$ depend on the coordinates $x^A$ only. The corresponding reduced Cartan forms \p{formsA} read
\bea\label{formsA1}
&& \left( \omega_P\right)^A_{red} =\cE^A{}_B dx^B,\quad \cE^A{}_B \equiv  \delta_B^A -\im \left( \psi^\alpha_a \partial_B \bpsi^{a \dot\alpha}
+ \bpsi^{a \dot\alpha} \partial_B  \psi^\alpha_a \right) \left( \sigma^A  \right)_{\alpha\dot\alpha}, \nn \\
&&\left(\omega_Z\right)_{red}^{ia} = d q^{ia}, \quad
(\omega_S)^{a \alpha}_{red}  =  d\psi^{a \alpha},\quad (\bar\omega_S)^{a\dot\alpha}_{red}  =  d\bpsi{}^{a\dot\alpha}.
\eea
These forms are invariant with respect to the transformations \p{susyS1}. Therefore, the covariant $x$-derivative will be
\be\label{cD}
\cD_A = \left(\cE^{-1}\right){}_A{}^B \partial_B,
\ee
while the invariant volume can be constructed from the forms $\left( \omega_P\right)^A_{red}$. Thus, the proper covariantization of the action
\p{bos_action}, having the right bosonic limit, will be
\be\label{actA}
S_1 = \int d^4 x \det (\cE)  \; \sqrt{- \cG},
\ee
where the covariantized metric tensor $\cG_{AB}$, evidently, reads
\be\label{G}
\cG_{AB} \equiv \eta_{AB} - 2\cD_{A} q^{ia}\cD_B q_{ia}\,, \quad \cG = \det \cG_{AB}\,.
\ee
The action $S_1$ \p{actA} reproduces the  fixed kinetic terms for bosons and fermions
\be
\left(S_1\right)_{lin} = -\int d^4 x \Big[ \im \left(  \psi_a^\alpha \partial_{\alpha\dot\alpha}\bpsi^{a \dot\alpha}
+ \bpsi^{a \dot\alpha} \partial_{\alpha\dot\alpha}\psi_a^\alpha  \right)
+ 2 \partial^A q^{ia} \partial_A q_{ia} \Big].
\ee
This would be too strong to maintain unbroken supersymmetry. Therefore, we have to introduce one more, evidently invariant, purely fermionic action
\be\label{actB}
S_2= \alpha \int d^4x \; \det (\cE),
\ee
which will correct the kinetic terms for the fermions, because
\be
\left(S_2\right)_{lin} = - \im \alpha \int d^4 x \left(  \psi_a^\alpha \partial_{\alpha\dot\alpha}\bpsi^{a \dot\alpha}
+ \bpsi^{a \dot\alpha} \partial_{\alpha\dot\alpha}\psi_a^\alpha  \right).
\ee
Thus, our Ansatz for the invariant supersymmetric action of the 3-brane acquires the form
\be\label{actAB}
S= S_0+S_1+S_2 = \left( 1+ \alpha\right) \int d^4x - \int d^4 x \det (\cE)\Big[ \alpha + \sqrt{- \cG}\, \Big],
\ee
where $\alpha$ is a constant parameter that has to be defined, and we have added the trivial invariant action $S_0 = \int d^4x$ to have a proper limit
$$ S_{q,\psi \rightarrow 0} =0.$$

Let us finish this Subsection by some comments on the Ansatz for our action \p{actAB}
\begin{itemize}
\item Firstly, the fermions $\psi_a^\alpha,\, \bpsi^{a\dot\alpha}$ transform under broken supersymmetry \p{susyS1} as the Goldstinos of the Volkov-Akulov model \cite{VA}.
This means that they may enter an invariant action only through the determinant of the super-vielbein $\det(\cE)$ or the covariant derivatives $\cD_A$. The action \p{actAB}
has just this structure. Moreover, the fermionic limit of the action
$$
S_{ferm} =\left( 1+ \alpha\right) \int d^4x \Big[ 1-  \det (\cE)\Big]
$$
is just the Volkov-Akulov action for the Goldstino, in full agreement with the results of \cite{sk1,sk2}.
\item Secondly, the transformation properties of the fields $q^{ia}$ \p{susyS1} show that $q^{ia}$ are just the matter fields with respect to broken supersymmetry.
Therefore, it is clear that any action of the form
$$
S= \int d^4x \det (\cE)\, \cF( \cD q)\,,
$$
where $\cF$ is an arbitrary function, depending on all possible  Lorentz and $SU(2)\times SU(2)$ invariant combinations constructed from $\cD_A q^{ia}$, invariant
under unbroken supersymmetry. Thus, the knowledge of the proper bosonic limit of the action \p{bos_action} is very important to select the particular system from the
quite wide family of the actions invariant with respect to broken supersymmetry.
\item Thirdly, the parameter $\alpha$ can be immediately fixed to be $\alpha=1$, if we will insist on the invariance of the linearized action
$\left(S_1\right)_{lin}+\left(S_2\right)_{lin}$
under linearized transformations of unbroken supersymmetry
$$\delta^{lin}_Q q^{ia} =2 \im \left( \epsilon^{i\alpha} \psi^a_\alpha +\bar\epsilon^{\,i\dot\alpha} \bpsi^a_{\dot\alpha}\right), \quad
\delta^{lin}_Q \psi^a_\alpha = -\bar\epsilon^{\,\dot\alpha}_i \partial_{\alpha\dot\alpha} q^{ia},\quad
\delta^{lin}_Q \bpsi^a_{\dot\alpha} =\epsilon^\alpha_i \partial_{\alpha\dot\alpha} q^{ia}.
$$
Thus, the suitable Ansatz for our action is
\be\label{actABc}
S= 2 \int d^4x - \int d^4 x \det (\cE)\Big[ 1 + \sqrt{- \cG}\, \Big].
\ee
\item Finally, it seems to be strange to call the action \p{actABc} by the term Ansatz, because it does not contain any free parameter.
The reason for such a nomenclature is
simple - the action \p{actABc} is not the most general action possessing the proper bosonic limit \p{bos_action} and invariant under broken supersymmetry due
to existence of Wess-Zumino terms. Thus, the proper Ansatz for the action of 3-brane in $D=8$ reads
\be\label{actABC}
S= 2 \int d^4x - \int d^4 x \det (\cE)\Big[ 1 + \sqrt{- \cG}\, \Big] +  S_{WZ}\,.
\ee
This additional term $S_{WZ}$ present in the our Ansatz \p{actABC}, will be constructed in the next Subsection.
\end{itemize}
\subsection{Wess-Zumino term}
The construction of the Wess-Zumino term, which is not strictly invariant, but which is shifted by a total derivative under broken
supersymmetry \p{susyS1}, goes in a standard way \cite{luca}. First of all, one has to determine the closed  five form $\Omega_5$,
which is invariant under $d=4$ Lorentz and broken supersymmetry transformations \p{susyS1}. Moreover, in the present case
this form has to disappear in the bosonic limit, because our Ansatz for the action \p{actAB} already reproduces the proper bosonic action of the 3-brane \p{bos_action}.
Such a form can be easily constructed in terms of the Cartan forms \p{formsA1} (for the sake of brevity, we have omitted the subscript {\it "red"} below)  :
\bea\label{wz1}
\Omega_5 &=& \omega_S{}^\alpha_a \wedge \bar\omega_S{}^{b\dot\alpha}\wedge \omega_Z{}^{ia} \wedge \omega_Z{}_{ib}
\wedge \omega_P^{A} \; \left(\sigma_A\right)_{\alpha\dot\alpha} \nn \\
&=& d\psi_a^\alpha \wedge d \bpsi{}^{b\dot\alpha} \wedge dq^{ia} \wedge dq_{ib} \wedge  \left( dx_{\alpha\dot\alpha}
-2 \im\left( \psi_{c\alpha}d\bpsi^c_{\dot\alpha}+\bpsi^c_{\dot\alpha} d \psi_{c\alpha}\right)\right).
\eea
To see that $\Omega_5$ \p{wz1} is indeed a closed form, one should take into account that the exterior derivative of $\left(\omega_P\right)_{\alpha\dot\alpha}$ is given
by the expression
\be\label{add1}
d \left(\omega_P\right)_{\alpha\dot\alpha} = -4 \im\, d\psi_{c\alpha} \wedge d\bpsi^c_{\dot\alpha}\,,
\ee
and, therefore, $d\Omega_5=0$, because
$$ d\psi_a^\alpha \wedge d\psi_{b\alpha} =\frac{1}{2}\, \epsilon_{ab} d\psi^{c\alpha} \wedge d\psi_{c\alpha} \;\Rightarrow \;
d \Omega_5 \sim d\psi^{a\alpha} \wedge d\psi_{a\alpha}\wedge d\bpsi^{b\dot\alpha} \wedge d\bpsi_{b\dot\alpha}\wedge dq^{ic} \wedge dq_{ic}=0 . $$
Next, one has to write $\Omega_5$ as the exterior derivative of a 4-form $\Omega_4$. This step, in contrast with the case of supersymmetric 3-brane in $D=6$ \cite{part1}, is not
completely trivial. Starting with the "evident" guess
\be\label{omega4}
\Omega_4^{(1)}=\frac{1}{2}\left( \psi_a^\alpha  d \bpsi{}^{b\dot\alpha}+\bpsi{}^{b\dot\alpha}d \psi_a^\alpha\right)
\wedge dq^{ia} \wedge dq_{ib} \wedge  \left( dx_{\alpha\dot\alpha} -2 \im\left( \psi_{c\alpha}d\bpsi^c_{\dot\alpha}+\bpsi^c_{\dot\alpha} d \psi_{c\alpha}\right)\right),
\ee
we will get
\be\label{step2}
d \Omega_4^{(1)} = \Omega_5+ 2 \im \left( \psi_a^\alpha  d \bpsi{}^{b\dot\alpha}+\bpsi{}^{b\dot\alpha}d \psi_a^\alpha\right)
\wedge d\psi_{c\alpha}\wedge d\bpsi^c_{\dot\alpha}\wedge dq^{ia}\wedge dq_{ib}.
\ee
The last step is to note that the second term in r.h.s of \p{step2} may be represented as
\be
- \im d \big\{ \left[ (\psi^{a\alpha}d\psi_\alpha^b) \wedge (\bpsi^{c\dot\alpha}\wedge d\bpsi_{c\dot\alpha})+ (\psi^{c\alpha}
\wedge d\psi_{c\alpha}) \wedge (\bpsi^{a\dot\alpha}d\bpsi^b_{\dot\alpha})\right]\wedge dq^i_a\wedge dq_{ib}\big\}.
\ee
Thus, the proper form $\Omega_4$, with the property $d\Omega_4=\Omega_5$, is given by the expression
\bea\label{wz2}
\Omega_4 &=& \frac{1}{2}\left( \psi_a^\alpha  d \bpsi{}^{b\dot\alpha}+\bpsi{}^{b\dot\alpha}d \psi_a^\alpha\right)
\wedge dq^{ia} \wedge dq_{ib} \wedge  \left( dx_{\alpha\dot\alpha} -2 \im\left( \psi_{c\alpha}d\bpsi^c_{\dot\alpha}
+\bpsi^c_{\dot\alpha} d \psi_{c\alpha}\right)\right)  \nn \\
&+&\im  \left[ (\psi^{a\alpha}d\psi_\alpha^b) \wedge (\bpsi^{c\dot\alpha}\wedge d\bpsi_{c\dot\alpha})+ (\psi^{c\alpha}
\wedge d\psi_{c\alpha}) \wedge (\bpsi^{a\dot\alpha}d\bpsi^b_{\dot\alpha})\right]\wedge dq^i_a\wedge dq_{ib}.
\eea
Integrating this form \p{wz2} we will get the Wess-Zumino action
\bea\label{WZ}
S_{WZ} &=& \beta \int d^4 x \det(\cE)\, \epsilon^{ABCD} \Big[
\left( \psi_a^\alpha  \cD_A \bpsi{}^{b\dot\alpha}+\bpsi{}^{b\dot\alpha}\cD_A \psi_a^\alpha\right) \cD_B q^{ia} \cD_C q_{ib}
\left( \sigma_D\right)_{\alpha\dot\alpha} \nn\\
&-&  2\im \left( \psi_a^{\alpha}\cD_A \psi_\alpha^b\; \bpsi^{c\dot\alpha}\cD_B \bpsi_{c\dot\alpha}+ \psi^{c\alpha}\cD_A \psi_{c\alpha}\;
\bpsi_a^{\dot\alpha}\cD_B \bpsi^b_{\dot\alpha}\right)\cD_C q^{ia}\cD_D q_{ib}\,\Big].
\eea
By construction, the action \p{WZ}  is invariant with respect to broken supersymmetry transformations \p{susyS1}.
Thus, the full Ansatz for the component action of 3-brane in $D=8$
reads
\bea\label{action}
S&=& 2 \int d^4x - \int d^4 x \det (\cE)\left[ 1 + \sqrt{- \cG}\right]   \nn \\
&+& \beta \int d^4 x \det(\cE)\, \epsilon^{ABCD} \Big[
\left( \psi_a^\alpha  \cD_A \bpsi{}^{b\dot\alpha}+\bpsi{}^{b\dot\alpha}\cD_A \psi_a^\alpha\right) \cD_B q^{ia}
\cD_C q_{ib}\left( \sigma_D\right)_{\alpha\dot\alpha} \nn\\
&-&  2\im \left( \psi_a^{\alpha}\cD_A \psi_\alpha^b\; \bpsi^{c\dot\alpha}\cD_B \bpsi_{c\dot\alpha}
+ \psi^{c\alpha}\cD_A \psi_{c\alpha}\;  \bpsi_a^{\dot\alpha}\cD_B \bpsi^b_{\dot\alpha}\right)\cD_C q^{ia}\cD_D q_{ib}\,\Big].
\eea
One should note that the Ansatz \p{action} is the unique, minimal, namely containing only the first derivatives
of the fields involved, action with the proper bosonic limit, which is invariant with respect to the broken supersymmetry \p{susyS1}.
The role of the unbroken supersymmetry is to fix the constant parameter $\beta$ (we already used the linearized version of the unbroken
supersymmetry to fix the parameter $\alpha$ in \p{actAB}).

\subsection{Unbroken supersymmetry}
The most technically complicated part of our approach is to maintain the unbroken supersymmetry, despite the fact that all we need is to fix one parameter in the action \p{action}.
In a case which we have now in hands, the situation is even worse than that was in the cases which we considered before \cite{part1, BIKS1, BKSu1, BIKS2, BIKS3},
because until now we did not present the
exact expressions for the quantities $(\mJ_A)^{ia}, (\mbJ_A)^{ia}$ \p{eq3} in terms of $\nabla_A \mq^{ia}$. The corresponding equations, which can be obtained by the action of
the anticommutators $\left\{ \nabla_{i \alpha}, \bnabla_{j \dot\alpha} \right\}$ \p{alg_der} on the superfield $\mq^{mb}$, have the form
\be\label{nleq1}
\delta_j^m (\mJ_{\alpha\dot\alpha})_i^b - \delta_i^m (\mbJ_{\alpha\dot\alpha})^b_j = \epsilon_{ij} \nabla_{\alpha\dot\alpha} \mq^{mb}
+ (\mJ_{\alpha\dot\gamma})^{a}_i (\mbJ_{\gamma\dot\alpha})_{aj} \nabla^{\gamma\dot\gamma} \mq^{mb}.
\ee
Passing to the components does not help, because the $\theta=\bar\theta=0$ projection of the equations \p{nleq1} reads quite similarly, to be
\be\label{nleq2}
\delta_j^m (J_{\alpha\dot\alpha})_i^b - \delta_i^m (\bJ_{\alpha\dot\alpha})^b_j = \epsilon_{ij} \cD_{\alpha\dot\alpha} q^{mb}
+ (J_{\alpha\dot\gamma})^{a}_i (\bJ_{\gamma\dot\alpha})_{aj} \cD^{\gamma\dot\gamma} q^{mb}.
\ee
It seems to be a completely hopeless idea to solve the equations \p{nleq2} starting from the most general Ansatz for $(J_A){}^{ia}$:
\bea\label{Janzatz}
(J_A)^{ia} &=& f_0\; \cD_A q^{ia} + f_1\; d_A^B\; \cD_B q^{ia} + f_2\; d_A^B \; d_B^C\;  \cD_C q^{ia}  \nn \\
&+& f_3 \; d_A^B\;  d_B^C\; d_C^D\; \cD_D q^{ia} +  f_4\; \epsilon_{ABCD}\; \cD^B q^{i b}\; \cD^C q_b^j\; \cD^D q_{j}^a,
\eea
where all functions $f$ are the complex(!) scalar functions depending, in general, on all possible Lorentz and $SU(2)\times SU(2)$ invariants
constructed from $\cD_A q^{ia}$, and
\be\label{d}
d_{AB} \equiv \cD_A q^{ia} \cD_{B} q_{ia}.
\ee
Indeed, it is completely unbelievable, that the  system of quadratically nonlinear equations for the five complex functions (if we will
succeed in their construction!) can be solved. Nevertheless, the iterative solution of the equations \p{nleq2} can be
straightforwardly found. In the first five orders in $\cD_A q^{ia}$ it reads
\bea\label{8dJomegaS}
(J_A)^{ia} &=& \cD_A q^{ia} +\Big\{ d_A^{B}\cD_B q^{ia} -\frac{1}{2}\, Tr \big(d\big) \, \cD_A q^{ia}
+ \frac{\im}{3}\,\epsilon_{ABCD} \cD^B q^{ib} \cD^C q^j_b \cD^D q_{j}^{a} \Big\}_3  \nn \\
&+& \Big\{ 2 d_A^{B}d_{BC}\cD^C q^{ia} - Tr \big(d\big) \, d_A^{B} \cD_B q^{ia} - Tr \big(d^2\big) \cD_A q^{ia}
+ \frac{1}{2}\, \big( Tr\big(d\big) \big)^2 \cD_A q^{ia}  \nn\\
&+& \frac{\im}{6} \big( \epsilon^{BCDE} \cD_B q^{j}_c \cD_C q_{j}^b \cD_D q_b^k \cD_E q_{k}^c \big)
\nabla_A q^{ia} \Big\}_5 + \Big\{ \ldots \Big\}_{\geq 7}\,.
\eea
The iterative solution \p{8dJomegaS} suggests another form of the $ (J_A)^{ia}$:
\be\label{JNK}
(J_A)^{ia} = N^B_A\, \cD_B q^{ia} + \im\, K^B_A\,(X_B)^{ia}\,,
\ee
where
\be\label{X}
(X^A)^{ia} =\epsilon^{ABCD} \cD_B q^{ib} \cD_C q^j_b \cD_D q_j^a
\ee
and the first entries in the {\it real} matrices functions $N_A^B$ and $K_A^B$ read
\bea\label{NK}
N^B_A &=& \Big[ 1- \frac{1}{2} Tr \big(d\big) + \frac{1}{2} \big( Tr \big (d \big ) \big )^2 - Tr \big ( d^2 \big )
- \frac{8}{3}\, Tr \big ( d^3 \big ) + \frac{11}{4}\, Tr \big( d^2 \big ) Tr \big( d \big ) - \frac{17}{24}\, \big( Tr \big (d \big ) \big )^3
\,\Big]\, \delta^B_A \nn\\
&+&\Big [1- Tr \big(d\big) + \frac{7}{4}\, \big( Tr \big (d \big ) \big )^2 - \frac{5}{2}\, Tr \big ( d^2 \big ) \,\Big]\, d_A^B
+ \Big [2 - 4\, Tr \big(d\big) \,\Big]\, \big (d \cdot d \big)_A^B + 6 \big( d \cdot d \cdot d \big) _A^B+\ldots \nn \\
K^B_A &=& \Big [\, \frac{1}{3} + \frac{1}{12}\, \big( Tr \big (d \big ) \big )^2 - \frac{1}{6}\, Tr \big( d^2 \big )\,
\Big ]\, \delta^B_A
-  \Big [\, \frac{2}{3} - \frac{1}{3} Tr \big(d\big)\, \Big ]\, d_A^B
- \frac{2}{3}\, \big (d \cdot d \big)_A^B+\ldots\,.
\eea
As we expected, the solution \p{NK} is rather complicated. Fortunately, there is a loophole which helps us to find the full
solution of the equations \p{nleq2} without solving them directly. The idea is to analyze the variation of our action
\p{action} with respect to unbroken supersymmetry keeping $(J_A)^{ia}$ arbitrary and then to find and solve the {\it linear}
equations which  guarantee the invariance of our action.
\subsubsection{Fixing $\beta$}
Before performing above mentioned analysis (see the next Subsection), let us firstly fix the parameter $\beta$.
In order to do this, one has expand the action \p{action} up to terms of the fourth order in $\partial_A q^{ia}$ and the second order in the fermions, i.e.
\bea\label{lim}
\det (\cE) & \Rightarrow & 1 - 2 \im \left( \psi^\alpha_a \partial_{\alpha\dot\alpha} \bpsi^{a \dot\alpha}
+ \bpsi^{a \dot\alpha}\partial_{\alpha\dot\alpha}  \psi^\alpha_a \right), \nn\\
1+\sqrt{- \cG}& \Rightarrow & 2 \Big( 1 - Tr \big(d\big) + \big(Tr \big(d\big) \big)^2 - Tr \big(d^2\big) \Big), \nn \\
S_{WZ} & \Rightarrow &
\beta \int d^4 x\,  \epsilon^{ABCD} \Big[
\left( \psi_a^\alpha  \partial_A \bpsi{}^{b\dot\alpha}+\bpsi{}^{b\dot\alpha}\partial_A \psi_a^\alpha\right)
\partial_B q^{ia} \partial_C q_{ib}\left( \sigma_D\right)_{\alpha\dot\alpha}\Big]
\eea
and consider the variation of this reduced action in the first order in the fermions and the third order in $\partial_A q^{ia}$.

Keeping in mind that under the unbroken $Q$ supersymmetry the covariant derivatives  $\nabla_A$ \p{4dsusy5}
are invariant, and then for an arbitrary superfield $\cal F$
$$
\left(\delta_Q \cF\right)_{\theta=\bar\theta=0}  = - \left(\epsilon^{\alpha}_i D^i_{\alpha} \cF
+\bar\epsilon^{\,i \dot\alpha} \bD_{i \dot\alpha} \cF\right)_{\theta=\bar\theta=0},\quad
\left(\delta_Q \nabla_A \cF\right)_{\theta=\bar\theta=0}  = - \left(\epsilon^{\alpha}_i D^i_{\alpha}\nabla_A \cF
+\bar\epsilon^{\,i \dot\alpha} \bD_{i \dot\alpha}\nabla_A\cF\right)_{\theta=\bar\theta=0},
$$
one may find the transformation properties of all ingredients in the action \p{action} (we will explicitly write only their $\epsilon$-part)
\bea\label{Qsusy-var}
\delta_{Q} \psi_{a \alpha} &=& H^B \partial_B \psi_{a \alpha}\,,\quad
\delta_{Q} \bar \psi^a_{\dot\alpha} = H^B \partial_B \bar \psi^a_{\dot\alpha} - \epsilon^{\alpha}_i (J_A)^{ia} \left(\sigma^A\right)_{\alpha\dot\alpha}\,, \\
\delta_{Q} \cD_A q_{ka} &=& H^B \partial_B \cD_A q_{ka} + 2i\, \epsilon^{\alpha}_k \cD_A \psi_{a \alpha}
- 2i\, \epsilon^{\alpha}_i \cD_A \psi_{b \alpha}\, (J^B)^{ib} \cD_B q_{ka} \nn\\
&-& 2\, \epsilon^{\alpha}_i \cD_A \psi_{b \beta}\, (J_D)^{ib} \big(\sigma^{DB} \big)_{\alpha}^{\;\;\;\beta} \cD_B q_{ka}\,, \nn\\
\delta_{Q} \cD_A \psi_{a \alpha} &=& H^B \partial_B \cD_A \psi_{a \alpha} - 2i\, \epsilon^{\beta}_i \cD_A \psi_{b \beta} (J^B)^{ib} \cD_B \psi_{a \alpha}
- 2\, \epsilon^{\gamma}_i \cD_A \psi_{b \beta} (J_D)^{ib} \big (\sigma^{DB} \big)_{\gamma}^{\;\;\;\beta} \cD_B \psi_{a \alpha}\,, \nn\\
\delta_{Q} \cD_A \bar \psi^a_{\dot\alpha} &=& H^B \partial_B \cD_A \bar \psi^a_{\dot\alpha}
- \epsilon^{\alpha}_i \cD_A (J_B)^{ia} \left(\sigma^B\right)_{\alpha\dot\alpha}
- 2i\, \epsilon^{\alpha}_i \cD_A \psi_{b \alpha}\, (J^B)^{ib} \cD_B \bar \psi^a_{\dot\alpha} \nn\\
&-& 2i\, \epsilon^{\alpha}_i \cD_A \psi_{b \beta}\, (J_D)^{ib} \big (\sigma^{DB} \big)_{\alpha}^{\;\;\;\beta} \cD_B \bar \psi^a_{\dot\alpha}\,, \nn
\eea
where
\be\label{def-H}
H^B = i \epsilon^{\alpha}_i \psi^a_{\alpha} (J^B)_a^{i} + \epsilon^{\alpha}_i \psi^a_{\beta} (J_A)_{a}^i\, \big(\sigma^{AB} \big)_{\alpha}^{\;\;\;\beta}.
\ee
As a consequence of \p{Qsusy-var} we will have
\bea\label{var-det}
\delta_{Q} \det ({\cal E}) &=& - \det ({\cal E})\, {\cal E}_B^{\;\;\;A}\, \delta_{Q} \big({\cal E}^{-1}\big)_A^{\;\;\;B} \nn \\
&=& \partial_A \big( H^A \det ({\cal E}) \big) + 2i\,\det ({\cal E})\, \epsilon^{\alpha}_i\, \Big [ \cD_A \psi_{a \alpha}\, (J^A)^{ia}
-i\, \cD_A \psi_{a \beta}\, (J_B)^{ia} \big(\sigma^{BA} \big)_{\alpha}^{\;\;\;\beta}\,
\Big].
\eea
Thus, we see that all $H$-dependent terms are converted into full space-time derivatives and, therefore, they can be ignored.
For the present analysis, in the above given
approximation we will need only the reduced version of these variations
\bea\label{Qsusy-var_red}
&&
\delta^{Red}_{Q} \psi_{a \alpha} = 0\,,\quad
\delta^{Red}_{Q} \bar \psi^a_{\dot\alpha} =  - \epsilon^{\alpha}_i (J_A)^{ia} \left(\sigma^A\right)_{\alpha\dot\alpha}\,, \\
&&
\delta^{Red}_{Q} \cD_A q_{ka} =  2i\, \epsilon^{\alpha}_k \partial_A \psi_{a \alpha}
- 2i\, \epsilon^{\alpha}_i \partial_A \psi_{b \alpha}\, (J^B)^{ib} \partial_B q_{ka}
- 2\, \epsilon^{\alpha}_i \partial_A \psi_{b \beta}\, (J_D)^{ib} \big(\sigma^{DB} \big)_{\alpha}^{\;\;\;\beta} \partial_B q_{ka}\,, \nn\\
&&
\delta^{Red}_{Q} \cD_A \psi_{a \alpha} = 0\,, \quad
\delta^{Red}_{Q} \cD_A \bar \psi^a_{\dot\alpha} =
- \epsilon^{\alpha}_i \cD_A (J_B)^{ia} \left(\sigma^B\right)_{\alpha\dot\alpha}\,, \nn \\
&& \delta^{Red}_{Q} \det ({\cal E}) =    2i\, \epsilon^{\alpha}_i\, \Big [ \partial_A \psi_{a \alpha}\, (J^A)^{ia}
-i\, \partial_A \psi_{a \beta}\, (J_B)^{ia} \big(\sigma^{BA} \big)_{\alpha}^{\;\;\;\beta}\,
\Big] \nn .
\eea
Moreover, in these transformations one should insert the object $(J_A)^{ia}$, up to the proper order, using the solution \p{8dJomegaS}.
Collecting all these together, we will get the following expression for the variation of the main part of the action \p{action}:
\bea\label{var11}
&&\delta^{Red}_Q\left[- \det (\cE) \left(1 + \sqrt{- \cG}  \right) \right] \approx
4\im \epsilon_{i\alpha} \partial_A \psi^\alpha_a \Big\{ \frac{\im}{3}\,\epsilon^{ABCD} \partial_B q^{ib} \partial_C q^j_b  \partial_D q^a_j \Big\}\nn \\
&& + 4 \epsilon_{i\alpha} \partial_A \psi^\beta_a \left(  \sigma^{AB} \right)^{\;\;\;\alpha}_\beta \Big\{\left(1 - Tr \big(d\big) \right) \partial_B q^{ia}
+ d_{BC}\partial^C q^{ia} + \frac{\im}{3} \epsilon_{BCDF} \partial^C q^{ib} \partial^D q^j_b  \partial^F q^a_j  \Big\}   \\
&&
+ 4 \epsilon_{i\alpha}\, \partial^C \psi^\beta_a \left( \sigma^{AB}  \right)_\beta^{\;\;\;\alpha} d_{AC} \partial_B q^{ia}.\nn
\eea
The term in the first line in r.h.s. of \p{var11}  cancels as a full divergence.  The terms containing both matrix and $\epsilon$-symbol  can be simplified
using the $\sigma$-matrices property
\bea\label{epsmatr}
\sigma^{BC}\epsilon_{CFGH} = \im \left( \delta^B_F \sigma_{GH} -\delta^B_G \sigma_{FH} +\delta^B_H \sigma_{FG}  \right).
\eea
The last term in \p{var11} can be also simplified by inserting $d_{CB} = \partial_C q^{ia} \partial_B q_{ia}$ and
extracting the terms antisymmetric in $\{i,j\}$ and $\{ a,b \}$, respectively. Then the term antisymmetric in $\{ a,b \}$, also cancels out.
Finally, the variation reads
\bea\label{var12}
\delta^{Red}_Q\left[ -\det (\cE) \left(1 + \sqrt{- \cG}  \right) \right] &\approx&  4 \epsilon_{i\alpha} \partial_A \psi^\beta_a \left(  \sigma^{AB} \right)^{\;\;\;\alpha}_\beta \Big\{\left(1 - Tr \big(d\big) \right) \partial_B q^{ia}
+ d_{BC}\partial^C q^{ia} \,\Big \}  \nn \\
&+&  4 \epsilon_{i\alpha} \partial^C \psi^\beta_a \left( \sigma^{AB} \right)^{\;\;\;\alpha}_\beta \partial_A q^{jb} \partial_C q^i_b \partial_B q^a_j\,.
\eea
The variation of the reduced Wess-Zumino action \p{lim} can be easily found, because in our approximation  we only need to vary $\bpsi^{i\dot\alpha}$, and take
 $(J_A)^{ia}$ in this variation only up to the lowest approximation. This variation involves the product of $\sigma$-matrices which can be simplified using  \p{epsmatr}.
Finally, we will get
\bea\label{var13}
\delta_Q {\cal L} _{WZ} &\approx& -2\beta \epsilon_{i\alpha} \left\{
\partial_C \psi^\beta_a \left( \sigma^{BC} \right)^{\;\;\;\alpha}_\beta \partial_B q^b_j \partial_A q^{ja} \partial^A q^i_b
+  \partial_C \psi^\beta_a  \left( \sigma^{AB} \right)^{\;\;\;\alpha}_\beta \partial^C q^i_b \partial_A q^{ja} \partial_B q^b_j \right. \nn\\
&-& \left. \partial_C \psi^\beta_a  \left( \sigma^{AC} \right)^{\;\;\;\alpha}_\beta \partial_A q^{ja} \partial_B q^b_j \partial^B q^i_b \right\}.
\eea
After a slightly rearranging of the terms, we will find that \p{var13} precisely cancels the variation \p{var12} if $\beta =2$. Thus, the action \p{action} with $\beta=2$ is
invariant under broken and unbroken supersymmetries in this approximation. One should note that, due to the fact that we do not have at hands any more freedom to modify the action,
and keeping in mind that the terms of higher order in the fermions come out from the lowest one, due to the invariance under broken supersymmetry, we have to conclude
that the full component action of 3-brane in $D=8$ reads
\bea\label{finaction}
S&=& 2 \int d^4x - \int d^4 x \det (\cE)\left[ 1 + \sqrt{- \cG}\right]   \nn \\
&+& 2 \int d^4 x \det(\cE)\, \epsilon^{ABCD} \Big[
\left( \psi_a^\alpha  \cD_A \bpsi{}^{b\dot\alpha}+\bpsi{}^{b\dot\alpha}\cD_A \psi_a^\alpha\right) \cD_B q^{ia} \cD_C q_{ib}\left( \sigma_D\right)_{\alpha\dot\alpha} \nn\\
&-&  2\im \left( \psi_a^{\alpha}\cD_A \psi_\alpha^b\; \bpsi^{c\dot\alpha}\cD_B \bpsi_{c\dot\alpha}+
\psi^{c\alpha}\cD_A \psi_{c\alpha}\;  \bpsi_a^{\dot\alpha}\cD_B \bpsi^b_{\dot\alpha}\right)\cD_C q^{ia}\cD_D q_{ib}\Big].
\eea
\subsubsection{Solution for $(J_A)^{ia}$}
The last step to complete our analysis of the unbroken supersymmetry is to find a closed expression for the $(J_A)^{ia}$ entering the transformation properties
\p{Qsusy-var}. Our attempts to solve the basic equations \p{nleq2} result only in the iterative solution \p{8dJomegaS} which may be prolonged up to any desired order, but which
cannot help us to find the full solution. The idea to find the full expression for $(J_A)^{ia}$ we shortly discussed above, is based on the invariance of our action \p{finaction}.
If this action is invariant, then the vanishing of its variation under transformations  \p{Qsusy-var} with unspecified $(J_A)^{ia}$ will result in the {\it linear}
equations on $(J_A)^{ia}$ which we are going to solve instead of solving the nonlinear equations \p{nleq2}. Good news is that the bosonic limit of the equations \p{nleq2}
simply corresponds to the replacement $\cD_A q^{ia}\rightarrow \partial_A q^{ia}$, and therefore it is enough to consider the variation of the action
\p{finaction} to the first order in the fermions.
If we will find the $(J_A)^{ia}$ which nullify this variation of the action, then the full $(J_A)^{ia}$ with all fermionic terms can be reconstructed from it by the inverse
substitution $\partial_A q^{ia} \rightarrow \cD_A q^{ia}$. The linear in the fermions variation of the integrand in the action \p{finaction} has the following form:
\bea\label{action-gen}
\delta_Q L &=& 4i\,  \sqrt{-g}\, \big( g^{-1}\big)^{AB}  \big( \Sigma_A \big)_{ia}\, \partial_B q^{ia}
- 2i\,  \Big [\delta_B^A + \sqrt{- g}\, \big( g^{-1}\big)_B^A \Big] \big( \Sigma_A \big)_{ia}\, (J^B)^{ia} \\
&-& 2\,  \Big [\delta_B^A + \sqrt{- g}\, \big( g^{-1} \big)_B^A \Big] \big( \Sigma_A^{\;\;\;DB} \big)_{ia}\, (J_D)^{ia} \nn\\
&-& 4 \, \epsilon^{ABCD} \epsilon^{\alpha}_k\, \partial_A \psi_{a \alpha}\, (J_D)^{kb} \partial_B q^{ia}\, \partial_C q_{ib}
+ 4 i \, \epsilon^{ABCD} \epsilon^{\alpha}_k\, \partial_A \psi_{a \beta}\, (J^F)^{kb}\, \big(\sigma_{FD} \big)_{\alpha}^{\;\beta}\,
\partial_B q^{ia}\, \partial_C q_{ib}\,, \nn
\eea
where we used the following notations:
\be\label{Sigma}
\big( \Sigma_A \big)_{ia} = \epsilon^{\alpha}_i\, \partial_A\, \psi_{a \alpha}\,,   \quad
\big( \Sigma_A^{\;\;\;DB} \big)_{ia} = \epsilon^{\alpha}_i\, \partial_A\, \psi_{a \beta}\, \big(\sigma^{DB} \big)_{\alpha}^{\;\;\;\beta}\,.
\ee
Combining together the terms with $\big( \Sigma_A \big)_{ia}$ and $\big( \Sigma_A^{\;\;\;DB} \big)_{ia}$ we will get
\bea
\delta_Q S & = & \int d^4x \big( \Sigma_A \big)_{ka} \Big\{
4 \im\, \sqrt{- g}\, \big( g^{-1}\big)^{AB}\, \partial_B q^{ka} - 2 \im\, \Big [\delta_B^A + \sqrt{- g}\, \big( g^{-1} \big)_B^A \Big]\, (J^B)^{ka}  \nn\\
&-&  4  \, \epsilon^{ABCD}\, \partial_B q^{ia}\, \partial_C q_{ib}\, (J_D)^{kb} \Big\} \label{var22a} \\
&+& 2\int d^4x \big( \Sigma_{A,DB} \big)_{ia}\left\{
\Big [\eta^{AB} + \sqrt{- g}\, \big( g^{-1} \big)^{AB} \Big]\, (J^D)^{ka}
- 2 \im \,  \epsilon^{ABCF}\, (J^D)^{kb}\, \partial_C q^{ia}\, \partial_F q_{ib}\right\}. \nn\\ \label{var22b}
\eea
Both variations, which are proportional to $\big( \Sigma_A \big)_{ia}$ and $\big( \Sigma_A^{\;\;\;DB} \big)_{ia}$, have to be zero independently. Nevertheless, due to a specific
structure of  $\big( \Sigma_A \big)_{ia}$ and $\big( \Sigma_A^{\;\;\;DB} \big)_{ia}$ \p{Sigma} we cannot conclude that the quantities in the curly brackets are equal to zero.
Indeed, one may easily check that the terms
\be\label{addterm}
\int d^4x  \big( \Sigma_A \big)_{ia} (X^A)^{ia} \quad\mbox{and}\quad \int d^4x \big( \Sigma_{A,DB} \big)_{ia} \eta^{AB} \partial^{D} q^{ia}
\ee
are equal to zero, being full space-time derivatives. Thus, the expressions in the curly brackets in \p{var22a} and \p{var22b}
have to vanish up to the integrand in these additional terms.
The coefficients before these terms can be easily fixed from the known lowest orders in the iterative solution \p{8dJomegaS}.
Thus, we came to the following equations:
\be\label{Eq1}
4 \im\, \sqrt{- g}\, \big( g^{-1}\big)^{AB}\, \partial_B q^{ka} - 2 \im\, M_B^A \, (J^B)^{ka}
- 2 \beta  \, \epsilon^{ABCD}\, \partial_B q^{ia}\, \partial_C q_{ib}\, (J_D)^{kb} = -\frac{8}{3} (X^A \big)^{ka}\,,
\ee
\be\label{Eq2}
2\, M^{A\underline{B}}\, J^{ka \underline{D}}
- 2 \im \beta\,  \epsilon^{A\underline{B}CF}\, (J^{\underline{D}})^{kb}\, \partial_C q^{ia}\, \partial_F q_{ib}
= 4 \eta^{A\underline{B}} \partial^{\underline{D}} q^{ka}\,,
\ee
where
\be\label{M}
M_B^A = \Big [\delta_B^A + \sqrt{- g}\, \big( g^{-1} \big)_B^A \Big]
\ee
and we underline the indices in the equation \p{Eq2} to remind that due to anti-symmetry and self-duality of $\sigma^{BD}$ in \p{Sigma} we have only three independent
equations over these indices in \p{Eq2}.

Let us start from the equation \p{Eq1}. In order to avoid the appearance of the $su(2)$ indices $(i,a)$,
we will convert this equation with $\partial_B q_{ka}$ and substitute the Ansatz \p{JNK}
for $(J_A)^{ia}$. Doing so, we finally get the following matrix equations on the real and imaginary parts of \p{Eq1}:
\bea\label{eqsKN}
&& \frac{1}{2}\, \big ( M \cdot K \big)^A_B - \frac{1}{3}\, \big (\,\delta^A_B\, \delta^C_D - \delta^A_D\, \delta^C_B \,  \big )\, N^D_C
= - \frac{2}{3}\, \delta^A_B\,, \nn \\
&& 4\, \sqrt{-g}\,\big ( g^{-1}\big )^{AB}\, d_{BD} - 2\, \big (\, M \cdot N \cdot d\, \big )^A_D
- \frac{4}{3}\, \big (\,\delta^A_D\, \delta^C_F - \delta^A_F\, \delta^C_D \,
\big )\, \big (\, K \cdot Z\, \big)_C^F = 0\,,
\eea
where the matrix $Z^{AB}$ reads
\be\label{ZZZ}
Z^{AB} = (X^A)^{ia} (X^B)_{ia}.
\ee
This matrix can be expressed through $d_{AB}$ using the definition \p{X}
\bea\label{X2X}
(X^A)^{ia} (X^B)_{ia} & = & -9/2 \left[ \left( \frac{2}{3} Tr(d^3)- Tr(d^2) Tr(d) +\frac{1}{3}(Tr(d))^3\right) \eta^{AB}+\left(Tr(d^2)-(Tr(d))^2\right) d^{AB}+\right. \nn \\
&& \left. 2 Tr(d)\; d^{AC}\; d_{C}^B -2 d^{AC}\;d_C^D\; d_D^B \right].
\eea
Besides the rather complicated structure of the matrix $Z^{AB}$, the equations \p{eqsKN} are linear and can be easily solved. The only problem is to represent the solution
in a readable form. We succeeded in the following form of the solution:
\bea
N^A_B &=&  \frac{2}{F}\,
\Big (\,
2\, \sqrt{-g} - 4\, Tr \big( d \big ) + 4\, \big (Tr \big( d \big ) \big )^2 - 4\, Tr \big( d^2 \big ) - \sqrt{-g}\, Tr\big( g^{-1} \big)\,
\Big )\, \delta^A_B \nn\\
&-& \frac{2\, \sqrt{-g}}{F}\, \Big (\,2 + \sqrt{- g}\, Tr\big( g^{-1} \big)\, \Big) \big ( g^{-1}\big )^A_B
+ \frac{8}{F}\, \Big ( 2\, \big (d \cdot d \big)^A_B\, + \big (1- 2\, Tr\big (d\big )\big )\, d^A_B
\,\Big),
\nn \\
K^A_B &=&   \frac{4}{3 F}\, \Big[ 2 \sqrt{- g} \left( g^{-1}\right)^A_B-2 g ^A_B -\sqrt{- g}\; Tr\big( g^{-1} \big)\delta^A_B \Big],
\label{solJJ}
\eea
where
\be\label{sol-F}
F = 8 + 8\, \sqrt{- g} -16\, Tr\big ( d \big) + 8\,\big (Tr \big( d \big ) \big )^2 - 8\, Tr \big( d^2 \big )
+ g\, \big (Tr\big( g^{-1} \big)\big)^2 - 4\, \sqrt{- g}\, Tr\big( g^{-1} \big)\,.
\ee
Funny enough, the matrices $N$ and $K$ are related in a quite simple way
\be\label{NKd}
\big (N \cdot d \big)^A_B = \gamma K^A_B + \frac{1}{2}\, \delta^A_B\,,
\ee
with
\be\label{gamma}
\gamma = - \frac{3}{8}\, \Big ( 2 - 2\, \sqrt{- g} + \sqrt{- g}\; Tr \big( g^{-1} \big) \Big )\,.
\ee

Having at hands the exact solution for $(J_A)^{ia}$ which guaranteed the invariance of the action \p{finaction} under unbroken supersymmetry, it is a matter of
long and rather complicated calculations to check that the equations \p{Eq2} and \p{nleq2} are satisfied. We did not find any simple way, besides the brute force
checking, to demonstrate this fact.

\setcounter{equation}0
\section{Conclusion}
In this second part of our paper we described the partial breaking of $N=1, D=8$ supersymmetry down to $N=2, d=4$ one within the nonlinear realization  approach.
The basic Goldstone superfield
associated with this breaking is the $N=2, d=4$ hypermultiplet $q^{ia}$ subjected to a nonlinear generalization of the standard hypermultiplet constraints \p{hyper}.
The dynamical equations which follow from these constraints are identified with the worldvolume supersymmetric equations of supersymmetric 3-brane in $D=8$. This part
of our paper is quite similar to the previously considered case of spontaneously broken $N=1, D=10$ supersymmetry down to $(1,0)\; d=6$ supersymmetry performed in \cite{bik2}.
After this superfield consideration, in the main part of this paper (Section3), we turned to the construction of the component on-shell action for this 3-brane. The building
blocks for the component action are the Cartan forms for the reduced coset \p{4dcoset} which are completely similar to the famous case considered by Volkov and Akulov \cite{VA}.
Thus, the first Ansatz for our action \p{actAB}, possessing the proper bosonic limit, can be constructed immediately. The first parameter $\alpha$ appearing on this stage
can be easily fixed to be 1 by the invariance under the simplest, linear part of the unbroken supersymmetry transformations. The existence of the Wess-Zumino term makes the
construction slightly more complicated. Fortunately, this additional term with the new parameter $\beta$ can be constructed from
the same Cartan forms in the way discussed in \cite{luca}. Thus, our final Ansatz for the action is defined uniquely \p{action}
up to one free parameter $\beta$. The fixing of this parameter is a more complicated task.
Luckily, it turns out that it is enough to consider the invariance of our Ansatz to the first nontrivial order in $\partial_A q^{ia}$.
Thus, we came to our main result - the component action of supersymmetric 3-brane in $D=8$ \p{finaction}.

In contrast with the 3-brane in $D=6$ \cite{part1}, in order to prove the invariance of our new action with respect to unbroken supersymmetry,
firstly  one has to solve the nonlinear matrix equation \p{nleq2}. This equation relates the spinor covariant derivatives of the spinor
superfields $\nabla^i_\alpha \bpsi_{a \dot\alpha}, \bnabla^i_{\dot\alpha} \psi_{a \alpha}$ and the space-time derivative of bosonic
superfields $\cD_{\alpha \dot\alpha} q^{ia}$. This equation has a nicely defined iterative solution \p{8dJomegaS}, but due to a rather
complicated most general Ansatz \p{Janzatz}, it includes five complex functions. In order to solve this problem we reversed the arguments
and wrote the variation of the action
\p{finaction} under unbroken supersymmetry, still keeping $(J_A)^{ia}$ unspecified. Then the vanishing of this variation results
in the {\it linear} equations on  $(J_A)^{ia}$,
which can be immediately solved. Funny enough, the correctness of the approach leads to the fact that this solution of the {\it linear} equations
solved simultaneously the nonlinear equations.

Thus, in these two papers we completed the construction of the component actions for all 3-branes from the famous brane scan of the paper
~\cite{branescan}. The constructed actions
contain only objects with clear meaning - vielbeins and covariant derivatives. The derivation of the Wess-Zumino term also needs only the Cartan forms on the reduced coset.
The quite simple form of the final action of the 3-brane in $D=8$ raised the question of the superfield formulations of such a system within harmonic \cite{HSS} or
projective \cite{PSS} superspaces. Being constructed, such a description would  make it possible to include into the game the $N=2, d=4$ matter
superfields, in such a way as to get
$N=4,d=4$  invariant systems due to the proper interaction with the hypermultiplet. Finally, one should mention
that $N=4,d=4 \rightarrow N=2,d=4$ partial breaking of supersymmetry can be achieved by using the $N=2, d=4$ vector
supermultiplet \cite{bik3,bik4,rev1} instead of of the hypermultiplet, which will result in the $N=2$ supersymmetric Born-Infeld action.
We are hoping the approach we are using here will help to solve this task.

\section*{Acknowledgments}
We are grateful to Dmitry Sorokin and Igor Bandos for valuable correspondence.

This work was partially supported by  the ERC Advanced Grant no. 226455 \textit{``Supersymmetry, Quantum Gravity and Gauge Fields''}~(\textit{SUPER\-FIELDS}).
The work of N.K. and S.K. was supported by RSCF grant 14-11-00598. The work of A.S. was partially supported by RFBR grant 13-02-9062 Arm-a.

\setcounter{equation}{0}
\def\theequation{A.\arabic{equation}}
\section*{Appendix: Superalgebra, coset space, transformations and Cartan forms}
In this Appendix we collected some formulas describing the nonlinear realization
of $N=1, D=8$ Poincar\'{e} group in its coset over its $N=1, d=4$ subgroup.

In $d=4$ notation the $N=1, D=8$  Poincar\'e superalgebra is a four central charges extended $N=4$ super-Poincar\'{e} algebra containing the following set of generators:
\be\label{1}
\mbox{ N=4, d=4 SUSY }\quad \propto \quad \left\{ P_A, Q^i_\alpha,\bQ_{i \dot\alpha}, S^a_\alpha,\bS_{a \dot\alpha}, Z^{i a}, L_{AB}, K^{i a}_{A}, T^{ij}, R^{ab} \right\}.
\ee
Here, $P_{A}$  and $Z^{i a}$ are $D=8$ translation generators,  $Q^i_{\alpha}, \bQ_{i \dot\alpha}$ and $S^a_\alpha, \bS_{a \dot\alpha}$
are the generators of super-translations, the generators $L_{AB}$ form $d=4$ Lorentz algebra $so(1,3)$, the generators $K^{ia}_{A}$
belong to the coset $SO(1,7)/SO(1,3)\times SU(2) \times SU(2)$, while the generators $T^{ij}$ and $R^{ab}$ span $su(2)\times su(2)$ subalgebra $(i,a=1,2)$.
The  commutation relations of $D=8$ Poincar\'{e} algebra in this basis read
\bea\label{D8Poincare}
&&\left[L_{AB}, L_{CD}  \right]\; = \;\im \left( - \eta_{AC} L_{BD} + \eta_{BC} L_{AD} - \eta_{BD} L_{AC} +\eta_{AD} L_{BC}   \right),  \nn \\
&&\left[ L_{AB}, P_{C}\right]\; = \; \im\left( - \eta_{AC}P_{B} + \eta_{BC} P_{A}\right), \; \left[ L_{AB}, K_{C}^{ia}\right] \;
= \; \im\left(- \eta_{AC}K_{B}^{ia} +\eta_{BC} K_{A}^{ia}\right), \nn \\
&&\left[ T^{ij}, T^{kl}   \right] \; = \; \im \left( \epsilon^{ik} T^{jl} +\epsilon^{jk} T^{il}+\epsilon^{il} T^{jk}+\epsilon^{jl} T^{ik}    \right), \nn \\
&&\left[ R^{ab}, R^{cd}   \right] \; = \; \im \left( \epsilon^{ac} R^{bd} +\epsilon^{bc} R^{ad}+\epsilon^{ad} R^{bc}+\epsilon^{bd} R^{ac}    \right), \nn \\
&&\left[ T^{ij}, K_{A}^{ka} \right] \; = \; \im \left( \epsilon^{ik}K_{A}^{ja} + \epsilon^{jk} K_{A}^{ia} \right), \; \left[ R^{ab}, K_{A}^{ic} \right]
\; = \; \im \left( \epsilon^{ac}K_{A}^{ib} + \epsilon^{bc} K_{A}^{ia} \right), \nn \\
&&\left[ T^{ij}, Z^{ka} \right] \; = \; \im \left( \epsilon^{ik}Z^{ja} + \epsilon^{jk} Z^{ia} \right), \; \left[ R^{ab}, Z^{ic} \right]
\; = \; \im \left( \epsilon^{ac}Z^{ib} + \epsilon^{bc} Z^{ia} \right), \nn \\
&&\left[ P_A, K_{B}^{ia} \right] \; = \; \im \eta_{AB}Z^{ia}, \; \left[ K_{A}^{ia}, Z^{jb}    \right] \; = \; - 2\im \epsilon^{ij} \epsilon^{ab}P_A, \nn \\
&&\left[ K_{A}^{ia}, K_B^{jb} \right] \; = \; 2\im \epsilon^{ij} \epsilon^{ab} L_{AB} -\im \eta_{AB} \left( \epsilon^{ab}  T^{ij}+\epsilon^{ij}  R^{ab}  \right) .
\eea
Here, $\eta_{AB}= \mbox{diag}(1,-1,-1,-1)$.

The eight supercharges $Q^i_\alpha, \; \bQ_{i \dot\alpha}, \; S^a_{\alpha}, \; \bS_{a \dot\alpha}$ obey the following (anti)commutation relations:
\bea\label{superD8Poincare}
&&\left\{ Q_{\alpha}^i,  \bQ_{\dot\alpha j}   \right\} \; = \; 2 \delta^i_j \left( \sigma^A \right)_{\alpha\dot\alpha} P_A, \;
\left\{ S_{\alpha}^a,  \bS_{\dot\alpha b}   \right\} \; = \; 2 \delta^a_b \left( \sigma^A \right)_{\alpha\dot\alpha} P_A, \nn \\
&&\left\{ Q_{\alpha}^i,  S_{\beta}^a   \right\} \; = \; 2 \epsilon_{\alpha\beta} Z^{ia},\;
\left\{  \bQ_{\dot\alpha i}, \bS_{\dot\beta a} \right\} \; = \; 2\epsilon_{\dot\alpha\dot\beta} Z_{ia};\nn \\
&&\left[ L_{AB}, Q_\alpha^i   \right] \; = \; -\frac{1}{2} \left( \sigma_{AB}  \right)_\alpha^\beta Q_\beta^i, \;
\left[ L_{AB}, \bQ_{\dot\alpha i}   \right] \; = \; \frac{1}{2} \left(\tilde\sigma_{AB}  \right)_{\dot\alpha}^{\dot\beta} \bQ_{\dot\beta i},     \nn \\
&&\left[ L_{AB}, S_\alpha^a   \right] \; = \; -\frac{1}{2} \left( \sigma_{AB}  \right)_\alpha^\beta S_\beta^a, \;
\left[ L_{AB}, \bS_{\dot\alpha a}   \right] \; =  \; \frac{1}{2} \left(\tilde\sigma_{AB}  \right)_{\dot\alpha}^{\dot\beta} \bS_{\dot\beta a};      \nn \\
&&\left[ K_A^{ia}, Q_\alpha^j    \right] \; = \; \im \left( \sigma_a  \right)_{\alpha\dot\alpha} \epsilon^{ij} \bS^{\dot\alpha a}, \;
\left[ K_A^{ia}, S_\alpha^b    \right] \; =  \; -\im \left( \sigma_A  \right)_{\alpha\dot\alpha}\epsilon^{ab} \bQ^{\dot\alpha i}, \nn \\
&&\left[ K_A^{ia}, \bQ_{\dot\alpha j}    \right]\; = \; \im \left( \sigma_A  \right)_{\alpha\dot\alpha}\delta^i_j S^{\alpha a}, \;
\left[ K_A^{ia}, \bS_{\dot\alpha b}    \right]\; =\; -\im \left( \sigma_a  \right)_{\alpha\dot\alpha} \delta^a_b Q^{\alpha i};  \nn \\
&&\left[ T^{ij}, Q_\alpha^k \right] \; = \; \im \left(  \epsilon^{ik}Q_\alpha^j + \epsilon^{jk}Q_\alpha^i    \right), \;
\left[ T^{ij}, \bQ_{\dot\alpha k} \right] \; = \; -\im \left(  \delta^{i}_{k} \bQ_{\dot\alpha}^j + \delta^{j}_{k} \bQ_{\dot\alpha}^i    \right),\nn \\
&& \left[ R^{ab}, S_{\alpha}^c\right] = \im \left(  \epsilon^{ac}S_\alpha^b + \epsilon^{bc}S_\alpha^a    \right), \;
\left[ R^{ab}, \bS_{\dot\alpha c }\right] = -\im \left( \delta^a_c \bS_{\dot\alpha}^b +\delta^b_c \bS_{\dot\alpha}^a  \right).
\eea
We define the coset element as follows:
\be\label{4dcosetA}
g = e^{\im x^A P_A} e^{\theta^\alpha_i Q^i_\alpha + \bar\theta^{i \dot\alpha}\bQ_{i \dot\alpha}} e^{\im \mq^{ia}Z_{ia}} e^{\mpsi^{\alpha}_a S^a_\alpha
+ \mbpsi^{a\dot\alpha} \bS_{a\dot\alpha}} e^{\im \mLambda^A_{ia} K_A^{ia}}.
\ee
Here, $\{ x^A, \theta_i^\alpha, \bar\theta{}^{i \dot\alpha} \}$ are $N=2, d=4$ superspace coordinates, while the remaining coset parameters are $N=2,d=4$ Goldstone superfields.
The local geometric properties of the system are specified by the Cartan forms
\bea\label{cartanA}
g^{-1}dg &=& \im \left(\omega_P \right)^A P_A  + \im \left(\omega_Z\right)^{ia} Z_{ia} +   \left( \omega_Q  \right)_i^\alpha Q^i_\alpha
+ \left( \bar\omega_Q  \right){}^{i \dot\alpha} \bQ_{i \dot\alpha} + \left( \omega_S  \right)_a^\alpha S^a_\alpha
+ \left( \bar\omega_S  \right){}^{ a\dot\alpha} \bS_{a \dot\alpha}+ \nn \\
&&  \im  \left( \omega_K  \right)_{i a}^A K^{ia}_A +\im \left( \bar\omega_K  \right)^A \bK_A + \im \left( \omega_L  \right)^{AB} L_{AB}.
\eea
In what follows, we will need the explicit expressions of the following forms:
\bea\label{formsA}
\left(\omega_P \right)^A &=& \triangle x^B \cosh \sqrt{2 \mLambda_B^{jb} \mLambda_{jb}^A}
-2 \triangle \mq^{jb} \mLambda^C_{jb} \frac{\sinh\sqrt{2\mLambda_C^{jb} \mLambda_{jb}^A}}{\sqrt{2\mLambda_C^{jb} \mLambda_{jb}^A}}, \nn \\
\left(\omega_Z\right)^{ia} &=& \triangle \mq^{jb} \cosh \sqrt{2 \mLambda_{jb}^A \mLambda_A^{ia}}
- \triangle x^A  \mLambda_{A}^{jb} \frac{\sinh\sqrt{2\mLambda^B_{jb} \mLambda^{ia}_B}}{\sqrt{2\mLambda^B_{jb} \mLambda^{ia}_B}}, \nn \\
\left( \omega_Q  \right)_i^\alpha &=&d\theta^\beta_j \left( \cosh\sqrt W   \right)_{i\beta}^{j\alpha}
+ d\mbpsi^{c\dot\gamma}\left(\frac{ \sinh \sqrt{\bT}}{\sqrt{\bT}}  \right)_{c\dot\gamma }^{b\dot\beta} \mLambda^\alpha_{ib\dot\beta}, \nn \\
\left( \omega_S  \right)_a^\alpha &=&d\mpsi^\beta_b \left( \cosh\sqrt{T}    \right)_{a\beta}^{b\alpha}
- d\bar\theta^{k\dot\gamma}\mLambda^\gamma_{bk\dot\gamma} \left( \frac{ \sinh \sqrt{T}}{\sqrt{T}} \right)_{a\gamma}^{b\alpha}  ,
\eea
where
\bea\label{forms_addA}
&&\triangle x^A = dx^A - \im \left( \theta^\alpha_i d\bar\theta^{i \dot\alpha} + \bar\theta^{i \dot\alpha}d \theta^\alpha_i
+ \mpsi^{\alpha}_a  d \mbpsi^{a\dot\alpha} + \mbpsi^{a\dot\alpha} d \mpsi^{\alpha}_a \right) \left( \sigma^A  \right)_{\alpha\dot\alpha}, \nn \\
&&\triangle \mq_{ia} = d \mq_{ia} -2\im \left( \mpsi_{a\alpha} d\theta^\alpha_i + \mbpsi_{a\dot\alpha} d\bar\theta^{\dot\alpha}_i     \right),
\eea
and the matrix-valued functions are defined as follows:
\be\label{mat_functions_A}
W_{i\beta}^{j\alpha} = \mLambda^{\alpha\dot\alpha}_{ia} \mLambda^{ja}_{\beta\dot\alpha}, \;
\bW_{j\dot\beta}^{i\dot\alpha} = \mLambda_{a\alpha}^{i\dot\alpha} \mLambda_{j\dot\beta}^{a\alpha},\quad T^{b\alpha}_{a \beta}
= \mLambda^{ib}_{\beta\dot\alpha} \mLambda^{\alpha\dot\alpha}_{ia}, \; \bT_{b\dot\beta}^{a\dot\alpha} = \mLambda^{i\alpha}_{b\dot\beta} \mLambda^{a\dot\alpha}_{i\alpha}.
\ee

Keeping in mind, that the  quantities $\triangle x^A, d\theta_i^\alpha$ and $d{\bar\theta}{}^{i \dot\alpha}$ are invariant with respect
to both supersymmetries, one may define the covariant derivatives $\nabla_A, \nabla^i_{\alpha}, \bnabla_{i \dot\alpha}$ as
\be\label{diff}
d \cF  = \left( dx^A \frac{\partial}{\partial x^A}+ d\theta_i^\alpha \frac{\partial}{\partial \theta_i^\alpha}+d\bar{\theta}^{i \dot\alpha}
\frac{\partial}{\partial \bar{\theta}^{i \dot\alpha}} \right) \cF = \left( \triangle x^A \nabla_A +d\theta_i^\alpha \nabla^i_\alpha
+ d{\bar\theta}{}^{i \dot\alpha} \bnabla_{i \dot\alpha}\right) \cF,
\ee
and, therefore,
\bea\label{4dsusy5}
\nabla_A &=& \left( E^{-1}  \right)_A^B \partial_B, \; E_A^B = \delta_A^B -\im \left( \mpsi^\alpha_a \partial_A \mbpsi^{a \dot\alpha}
+ \mbpsi^{a \dot\alpha} \partial_A  \mpsi^\alpha_a \right) \left( \sigma^B  \right)_{\alpha\dot\alpha},  \\
\nabla_\alpha^i  &=& D_\alpha^i -\im   \left( \mpsi^\beta_a \nabla_\alpha^i \mbpsi^{a \dot\beta}
+ \mbpsi^{a \dot\beta} \nabla_\alpha^i \mpsi^\beta_a \right) \left( \sigma^B  \right)_{\beta\dot\beta} \partial_B, \nn \\
\bnabla_{i \dot\alpha}  &=& \bD_{i \dot\alpha} -\im   \left( \mpsi^\beta_a \bnabla_{i \dot\alpha} \mbpsi^{a \dot\beta}
+ \mbpsi^{a \dot\beta} \bnabla_{i \dot\alpha} \mpsi^\beta_a \right) \left( \sigma^B  \right)_{\beta\dot\beta} \partial_B. \nn
\eea
Here, $D_\alpha^i, \; \bD_{i \dot\alpha}$ are flat derivatives obeying the relations
\be
\left\{ D_\alpha^i, \; \bD_{j \dot\alpha}    \right\} = -2\im \delta^i_j \left( \sigma^A  \right)_{\alpha\dot\alpha} \partial_A, \quad
\left\{ D_\alpha^i, \; D_\beta^j \right\}=\left\{ \bD_{i \dot\alpha}, \; \bD_{j \dot\beta}    \right\}=0.
\ee
The covariant derivatives \p{4dsusy5} satisfy the following (anti)commutation relations:
\bea\label{alg_der}
&&\left\{  \nabla_\alpha^i , \; \nabla_{\beta}^j   \right\} =  - 2\im \left(  \nabla_\alpha^i \mpsi_a^\gamma \nabla^j_\beta \mbpsi^{a\dot\gamma}
+  \nabla_\alpha^i \mbpsi^{a\dot\gamma} \nabla^j_\beta\mpsi_a^\gamma   \right) \left( \sigma^A   \right)_{\gamma\dot\gamma}\nabla_A, \nn \\
&&\left[  \nabla_A , \; \nabla_\alpha^i   \right] =  - 2\im \left(  \nabla_A \mpsi_a^\gamma \nabla_\alpha^i \mbpsi^{a\dot\gamma}
+  \nabla_A \mbpsi^{a\dot\gamma} \nabla_\alpha^i \mpsi_a^\gamma   \right) \left( \sigma^B   \right)_{\gamma\dot\gamma}\nabla_B, \nn \\
&&\left\{  \nabla_\alpha^i , \; \bnabla_{\dot\alpha j}   \right\} = -2\im \delta^i_j \left( \sigma^A  \right)_{\alpha\dot\alpha} \nabla_A
- 2\im \left(  \nabla_\alpha^i \mpsi_a^\gamma \bnabla_{j\dot\alpha } \mbpsi^{a\dot\gamma}
+  \nabla_\alpha^i \mbpsi^{a\dot\gamma} \bnabla_{j\dot\alpha }\mpsi_a^\gamma   \right) \left( \sigma^A   \right)_{\gamma\dot\gamma}\nabla_A.
\eea

\end{document}